\documentstyle[twocolumn,eqsecnum,aps]{revtex}

\begin{document}
\draft
\preprint{}
\title{Delta-Function Bose Gas Picture of $S=1$ Antiferromagnetic 
Quantum Spin Chains Near Critical Fields
}
\author{Kouichi Okunishi, Yasuhiro Hieida and Yasuhiro Akutsu}
\address{
Department of Physics, Graduate School of Science, Osaka University,\\
Machikaneyama-cho 1-1, Toyonaka, Osaka 560-0043, Japan.
}
\date{\today}
\maketitle
\begin{abstract}
We study the zero-temperature magnetization curve ($M-H$ curve) of the
one-dimensional quantum antiferromagnet of spin one. The Hamiltonian $H$ 
we consider is of the bilinear-biquadratic form:
$H=\sum_{i}f(\vec{s}_{i}\cdot\vec{s}_{i+1})$ (+Zeeman term) 
where $\vec{s}_{i}$ is the spin operator at site $i$ 
and $f(X)=X+{\beta}X^2$ with $0\leq{\beta}<1$. 
We focus on validity of the $\delta$-function bose-gas picture 
near the two critical fields: 
upper critical field $H_{s}$ above which the magnetization saturates 
and the lower critical field $H_{c}$ associated with the Haldane gap. 

As for the behavior near $H_{s}$, we take ``low-energy effective S-matrix'' 
approach where {\em correct} effective bose-gas coupling constant $c$ is 
extracted from the two-down-spin $S$-matrix in its low-energy limit. 
We find that the resulting value of $c$ differs from the spin-wave value. 
We draw the $M-H$ curve by using the resultant bose gas, 
and compare it with numerical calculation where 
the product-wavefunction renormalization group (PWFRG) method, 
a variant of the S. R. White's density-matrix renormalization group method, 
is employed. 
Excellent agreement is seen between the PWFRG calculation and the 
correctly-mapped bose gas calculation.

We also test the validity of the bose-gas picture 
near the lower critical field $H_{c}$. 
Comparing the PWFRG-calculated $M-H$ curves with the bose-gas prediction, 
we find that there are two distinct regions, 
I and II, of ${\beta}$ separated by a critical value ${\beta}_{c}$ 
($\approx 0.41$). 
In the region I, $0<{\beta}<{\beta}_{c}$, 
the effective bose coupling $c$ is positive but rather small. 
The small value of $c$ makes the ``critical region'' 
of the square-root behavior $M\sim \sqrt{H-H_{c}}$ very narrow. 
Further, we find that in the ${\beta} \rightarrow {\beta}_{c}-0$, 
the square-root behavior transmutes to a different one, 
$M\sim (H-H_{c})^{\theta}$ with $\theta\approx1/4$. 
In the region II, ${\beta}_{c}<{\beta} <1$, 
the square-root behavior is more pronounced as compared with the region I, 
but the effective coupling $c$ becomes {\em negative}.
\end{abstract}

\narrowtext

\section{Introduction}
\label{sec1}

The magnetization process ($M-H$ curve, $M$: magnetization, 
$H$: magnetic field) of one-dimensional (1D) 
quantum spin system has recently drawn much attention,  
due to the remarkable progress in material synthesis techniques 
and high-field experiments.~\cite{experiments}  
Spin chains with various $S$ (spin magnitude) 
and/or with non-trivial special structures, etc, 
exhibit various interesting magnetic behaviors 
(e.g., field-induced phase transitions, like plateau 
in the $M-H$ curve~\cite{M-H-plateau}), 
many of which still await theoretical analyses.

As for the $M-H$ curve of the gapful $S=1$ antiferromagnetic 
(AF) Heisenberg chain, it has been known that, 
on raising the magnetic field from zero, 
there is a  critical field $H_{c}$ 
above which the system becomes magnetized. 
This critical field relates to the excitation gap 
(Haldane gap) $\Delta$ as $H_{c}=\Delta/(g\mu_{\rm B})$ 
(, $g$: $g$-factor, $\mu_{\rm B}$: Bohr magneton).  
We call $H_{c}$ {\em lower critical field} 
because there also exists {\em upper critical field} $H_{s}$ 
(saturation field) above which the magnetization saturates to $M_{s}=1$. 
Near $H_{s}$, it is well established 
that the $M-H$ curve behaves as $M_{s}-M\sim\sqrt{H_{s}-H}$.
\cite{Parkinson-Bonner,Hodgson-Parkinson,Kiwata-Akutsu} 
Near $H_{c}$, similar behavior $M\sim(H-H_{c})^{\theta}$ 
with $\theta\approx 1/2$ has also been known,\cite{Affleck,Takahashi-Sakai} 
but numerically, whether the exponent $\theta$ is {\em exactly} 1/2 or not, 
has remained to be less conclusive as compared with the behavior near $H_{s}$.

  The expected square-root behavior $M\sim\sqrt{H-H_{c}}$ 
has been explained via approximate mapping to 
the $\delta$-function bose gas~\cite{Affleck,Takahashi-Sakai} 
or to the fermion gas.~\cite{Tsvelick}.  
There is numerical evidence for such
 mappings,~\cite{Takahashi-Sakai,Sorensen-Affleck} but, 
the square-root behavior of the {\em bulk magnetization} 
itself has not been fully verified yet.  
In the finite-$N$ ($N$: system size) diagonalization study,
\cite{Takahashi-Sakai}
the smallness of $N$ disables us to make quantitative discussion of 
the bulk magnetization near $H_{c}$.  
In Ref.~\cite{Sorensen-Affleck}, a large-size system is treated by the 
density-matrix renormalization group\cite{DMRG} (DMRG, for short) 
to study the low-lying excitations 
having small $S^{z}_{{\rm tot}}$ ($\leq 4$). 
It was shown there that the low-lying excitations admit fermionic 
interpretation, 
just as in the case of the Bethe-ansatz solution of 
the $\delta$-function bose gas at low particle density. 
This result is indeed a strong support for the bose-gas picture, 
but the smallness of $S^{z}_{{\rm tot}}$ implies that, in a strict sense, 
the result applies only to the system in the vanishing magnetization 
density $m=S^{z}_{{\rm tot}}/N \rightarrow0$. 

Further, there is a quantity which is undoubtedly important for the bose-gas 
picture, but has not been considered seriously so far: 
the effective bose-gas coupling constant $c$.  
Let us consider a 1D (effective) $\delta$-function bose 
gas with the Hamiltonian
\begin{equation}
{\cal H }^{\rm (bose)}(\sigma,c)= \int {\rm d}x [\sigma 
\partial\phi^{\dagger}(x) \partial \phi(x) +c 
\phi^{\dagger}(x)\phi^{\dagger}(x)\phi(x)\phi(x)], \label{tag-1}
\end{equation}
with $\phi^{\dagger}(x)$ and $\phi(x)$ being the bose operators.  
By solving the Bethe ansatz integral equation 
we obtain the ground-state energy density 
$\epsilon$ as\cite{Lieb-Liniger}
\begin{eqnarray}
\epsilon&=&\frac{c^3}{\sigma^2} \tilde{\epsilon}(r),\\
r&=& \frac{2\sigma\rho}{c}\quad\mbox{$\rho$: particle number density},\\
\tilde{\epsilon}(r)&=& \frac{\pi^2}{24} r^3
\left\{ 1-2r+3r^2+(\frac{4\pi^2}{15}-4)r^3 \right.\nonumber\\
&&\left.
+ (5- \frac{4\pi^2}{3} )r^4 +{\cal O}(r^5) \right\} .\label{tag-2}
\end{eqnarray}
In terms of the particle density $\rho$, we have
\begin{equation}
\epsilon(\rho)=A_{3}\rho^3 + A_{4}\rho^4 + \cdots,\label{tag-3}
\end{equation}
with
\begin{eqnarray}
A_{3}&=&\sigma\pi^2/3 \nonumber\\
A_{4}&=&-4\sigma^2\pi^2/(3c). \label{tag-4}
\end{eqnarray}
In the bose-gas picture, the $M-H$ curve of the spin chain near the critical 
field corresponds to the $\rho-\mu$ curve of the bose gas, 
where $\mu=\partial \epsilon(\rho)/\partial \rho$ is the chemical potential.  
As far as the critical behavior is concerned, 
only the $\rho^{3}$-term is relevant, which 
determines both the critical exponent ($=1/2$) and the critical amplitude. 
Since the coefficient $A_{3}$ does not depend on $c$, the critical behavior 
must be $c$-independent and ``universal''.  However, actual $\rho-\mu$ curve 
heavily dependents on the value of $c$ which comes from $\rho^4$ and/or 
higher-order terms.  In fact, the coefficient $A_{k}$ of $\rho^{k}$-term 
($k\geq 4$) is proportional to $c^{3-k}$, 
which becomes dominant for small $c$. 
Accordingly, the critical region of the square-root behavior will become rather
narrow for small $c$. Hence, knowledge of the actual value of the effective 
coupling constant $c$ is indispensable, in order to make a fully quantitative 
test of the bose-gas picture.

 The aim of the present article is to give a quantitative test for the bose-gas
picture of the bilinear-biquadratic AF chain in a field.  
For this purpose, we employ the quantum version~\cite{HOA} 
of the product-wavefunction renormalization group method\cite{PWFRG} 
(PWFRG method, for short). 
The PWFRG is a variant of the S.R. White's DMRG,~\cite{DMRG} 
which is specially designed to obtain the ``fixed point'' 
(=thermodynamic limit of the system~\cite{Ost-Rom}) 
of the DMRG iterations efficiently. In Ref.~\cite{HOA}, 
it is shown that the PWFRG which was originally implemented for 2D classical 
systems, can also be applied to 1D quantum systems by replacing the 
transfer-matrix multiplication with the modified Lanczos operation. Even with 
relatively small number of retained bases, which is conventionally denoted as 
``$m$'', PWFRG calculations accurately reproduce exact $M-H$ curves for 
integrable models,\cite{HOA,Sato-Akutsu} 
which demonstrates both the efficiency and the reliability of the method.

  The hamiltonian of the model we consider is
\begin{eqnarray}
{\cal H} &=& {\cal H}_{\rm BLBQ} - {\cal H}_{\rm zeeman}, \\
{\cal H}_{\rm BLBQ}&=&\sum_{i}\left[\vec{S}_{i}\cdot\vec{S}_{i+1} + 
{\beta}(\vec{S}_{i}\cdot\vec{S}_{i+1})^2\right] 
,\quad {\cal H}_{\rm zeeman} = H\sum_{i}S_{i}^{z}, 
\label{tag-5}
\end{eqnarray}
where $\vec{S}_{i}=(S_{i}^{x},S_{i}^{y},S_{i}^{z})$ is the $S=1$ spin operator 
at the site $i$. For notational simplicity, we have took the units where 
$g\mu_{B}=1$, or, these factors are absorbed into the field $H$. In this paper 
we consider the case $0\le{\beta}<1$.  We concentrate on the behavior of the 
$M-H$ curve near the two critical fields $H_{s}$ and $H_{c}$.

\section{Near the Saturation Field $H_{s}$}
Let us consider first the system near the upper critical field (saturation 
field) $H_{s}=4$, where the $M-H$ curve shows the square-root behavior: 
$1-M\sim \sqrt{H_{s}-H}$. 
As compared with the one near $H_{c}$, this behavior 
itself is well-established (by exact diagonalization~\cite{Parkinson-Bonner} 
and Bethe ansatz\cite{Hodgson-Parkinson,Kiwata-Akutsu}). 
Our concern is how well the $\delta$-function bose gas can describe 
the $M-H$ curve {\em away from} the saturation field.
For this purpose, we should derive the {\em correct} effective bose gas
hamiltonian, which we shall make by employing the 
low-energy effective $S$-matrix approach.

Above the saturation field $H_{s}$, the system is ferromagnetically ordered 
(``all-up'' state). 
Low-energy excitations slightly below $H_{s}$ are well 
described in terms of ``down spins'' in the sea of up (``$+1$'') spins.  
Regarding a down spin as a particle, 
we consider two-body scattering problem to 
obtain the exact two-body $S$-matrix.~\cite{Hodgson-Parkinson,Kiwata-Akutsu}
The point is that in the {\em low-energy limit} the $S$-matrix reduces, 
in most cases, to that of the $\delta$-function bose gas 
with an effective coupling 
constant. With this {\em correct} coupling constant, the bose gas will give a 
quantitative description of the system near $H_{s}$.

By $\{|\sigma_{1},\sigma_{2},\ldots,>\}$ ($\sigma_{i}=0,\pm1$), we denote the 
$S^{z}$-diagonal bases of the spin chain.  To solve the two-down-spin problem 
we express the eigenvector $|k,k'>$ in terms of the wavefunctions 
$\psi(x,y,k,k')$ and $f(z,k+k')$ as
\begin{eqnarray}
|k,k'>&=&\sum_{x<y}\psi(x,y,k,k')|11\cdots1\mathop{0}_{x} 
1\cdots1\mathop{0}_{y}1\cdots>\\
 &+& \sum_{z}f(z,k+k')|1\cdots1\mathop{(-1)}_{z}1\cdots>.\label{tag-6}
\end{eqnarray}
The two-body $S$-matrix $S(k,k')$ is introduced through the asymptotic 
($y-x\rightarrow\infty$) behavior of the wavefunction:
\begin{equation}
\psi(x,y,k,k') \sim e^{ikx}e^{ik'y} + S(k,k')e^{ik'x}e^{iky}. \label{tag-7}
\end{equation}
The eigenvalue problem ${\cal H}_{\rm BLBQ}|k,k'>=E(k,k')|k,k'>$ is solved to give
\begin{equation}
E(k,k')=\epsilon(k)+\epsilon(k') \label{tag-8}
\end{equation}
where $\epsilon(k)=-2+2\cos k$ is the one-particle (one-down-spin) energy. 
Since the one-particle energy $\epsilon(k)$ takes its minimum at $k=\pi$, we 
put
\begin{equation}
k=\pi+\kappa, k'=\pi+\kappa' \label{tag-9}
\end{equation}
and consider the ``low energy limit'' $\kappa,\kappa'\rightarrow0$. In this 
limit, the $S$-matrix whose explicit expression has been given in 
Ref.~\cite{Kiwata-Akutsu}, becomes
\begin{equation}
S(k,k') \mathop{\rightarrow}_{\kappa,\kappa \sim 0} S^{\rm (bose)}
(\kappa-\kappa',-\frac{3{\beta}+1}{{\beta}}) \label{tag-10}
\end{equation}
where $S^{\rm (bose)}(k,c)$ is the $S$-matrix of 
the $\delta$-function bose gas 
with hamiltonian ${\cal H }^{\rm (bose)}(1,c)$ in (\ref{tag-1}). 
Explicitly, we have
\begin{equation}
S^{\rm (bose)}(\kappa,c)=-\frac{c+i\kappa}{c-i\kappa}. \label{tag-12}
\end{equation}
Therefore, in discussing the low-energy properties near the saturation field, 
the bilinear-biquadratic chain (\ref{tag-5}) is equivalent 
to the bose-gas with the effective coupling constant
\begin{equation}
c=-\frac{3{\beta}+1}{{\beta}}. \label{tag-13}
\end{equation}
A remark is in order.  For ${\beta}>0$ or ${\beta}<-1/3$ the coupling constant 
(\ref{tag-13}) takes negative value.  It has been known that the 
$\delta$-function bose gas with negative coupling constant is unphysical, 
because the system is unstable against formation 
of multiparticle bound states.  
In the present case of finite-$S$ spin chain,
however, $N$-particle bound state with large $N$ are kinematically
forbidden because more than $2S$ ``particles'' 
cannot exist at a single site.  Further, in the $N=2$ case, the bound state 
actually exists,~\cite{Kiwata-Akutsu} but, for $0<{\beta}<1$ it is a 
high-energy mode which can also be neglected, near the saturation field at 
least.  Hence, the effective bose gas with negative coupling constant, does 
have a meaning in the present case.

From the bose-gas energy density $\epsilon(\rho)$ in (\ref{tag-2}) and 
(\ref{tag-3}) with $\sigma=1$ and $c=-(3\beta+1)/\beta$, we obtain the $M-H$ 
curve through
\begin{eqnarray}
M=1-\rho,\\
H_{s}-H=\frac{\partial\epsilon(\rho)}{\partial\rho}. \label{tag-14}
\end{eqnarray}
At arbitrary $\rho$, we can numerically solve the Bethe-ansatz integral 
equation~\cite{Lieb-Liniger} by converting it to a matrix equation, to obtain 
$\partial\epsilon(\rho)/\partial\rho$ within any required precision.  To check 
the validity of the bose-gas description of the $M-H$ curve, we performed 
the numerical-renormalization-group calculations. 
The method we employed is the 
quantum version~\cite{HOA} of the PWFRG,~\cite{PWFRG} which allows us to make 
fixed-$H$ calculations sweeping the value of $H$ giving the $M-H$ curve 
$M=M(H)$ in the thermodynamic limit.  
Fig.1 shows comparison between the bose-gas results 
and the PWFRG calculations, where we see excellent agreements 
for unexpectedly wide range of the field $H$. The validity of the bose-gas 
picture in the quantitative description of the $M-H$ curve, is thus verified.

 Our two-down-spin $S$-matrix approach is easily extended to general-$S$ 
bilinear-biquadratic chain with the hamiltonian
\begin{equation}
{\cal H} = \frac{1}{S}\sum_{i}\left[\vec{S}_{i}\cdot\vec{S}_{i+1} 
+ {\beta}(\vec{S}_{i}\cdot\vec{S}_{i+1})^2\right]. \label{generalS}
\end{equation}
After a straightforward calculation very similar to the ones in 
Refs.\cite{Hodgson-Parkinson,Kiwata-Akutsu}, we obtain the effective bose-gas 
coupling
as
\begin{equation}
c=-2\frac{1+{\beta}(3-8S+8S^2)}{1-S+{\beta}(3-9S+8S^2)} . \label{tag-15-1}
\end{equation}
For the pure bilinear case (${\beta}=0$), we have
\begin{equation}
c=2/(S-1) . \label{tag-15}
\end{equation}
In Fig.2 we compare the PWFRG-calculated curves for $S=1/2,1,3/2,2$ 
to bose-gas curves with corresponding values of $c$ given by (\ref{tag-15}).  
We again see satisfactory agreements. 

Let us give a comment on previous studies related to the present one.  For the 
spin-$S$ ``pure'' Heisenberg AF ($\beta=0$ in (\ref{generalS})), the bose-gas 
description near $H_{s}$ has already been made within the conventional 
spin-wave-theoretical approach.\cite{Johnson-Fowler,Takahashi-Sakai}
This approach gives the value of the bose-gas coupling constant to be
\begin{equation}
c=2/S ,\label{tag-15-0} 
\end{equation}
which is different from (\ref{tag-15}). 
The spin-wave value (\ref{tag-15-0}) deviates from (\ref{tag-15}) very much at 
small $S$ (even the {\em sign} disagrees at $S=1/2$), although both are the 
same in the large-$S$ limit.  
Having seen that the bose-gas with (\ref{tag-15}) gives the correct $M-H$ 
curve, we must say that, at small $S$, the spin-wave approach is not reliable 
enough for quantitative studies of the AF chains, even in the neighborhood of 
the saturation field. 
Note that, using the the Dyson-Maleev transformation and taking the continuum 
limit, we can formally rewrite\cite{Takahashi-Sakai} 
the spin-chain Hamiltonian 
(\ref{generalS}) with $\beta=0$ into the $\delta$-function bose-gas 
Hamiltonian with (\ref{tag-15-0}).  
Although this transformation seems to be exact in the operator level, there is 
a constraint on the state space: the boson number cannot exceed $2S$ at each 
site. 
This constraint amounts to ``kinematical interaction'' between the spin waves, 
which may be the source of the disagreement between (\ref{tag-15}) and 
(\ref{tag-15-0}).

\narrowtext
\section{Near the Lower Critical Field $H_{c}$}
At $H=0$ the ground state is singlet and non-magnetic.  
On raising $H$, system still remains to be singlet upto a critical field 
$H_{c}$ above which the ground-state become magnetized. 
The field-induced phase transition at $H_{c}$ is a level-crossing transition 
between the singlet state and the lowest-energy triplet state (both at $H=0$), 
hence the critical field $H_{c}$ is, in our unit, just the excitation gap 
(``Haldane gap'') $\Delta$.  
Then, in the bose-gas description near $H_{c}$, 
the singlet ground state should 
be interpreted as the ``vacuum'', and the triplet state with 
$S^{z}_{\rm tot}=1$ the ``one-particle state''.

 For ${\beta}\approx 0$, the ``one-particle'' energy dispersion $\omega(k)$ 
takes its minimum at $k=\pi$. 
The dispersion curve around this minimum is often assumed to be relativistic 
one,\cite{dispersion,Affleck}
\begin{equation}
\omega(k)=\sqrt{\Delta^2 + v^2\bar{k}^2}, \label{tag-16}
\end{equation}
where $\bar{k}=k-\pi$ and $v$ is called spin-wave velocity.  
In the low-energy ($|\bar{k}|\rightarrow0$) limit, (\ref{tag-16}) becomes
\begin{equation}
\omega(k)=\Delta + \frac{v^2}{2\Delta}\bar{k}^2. \label{tag-17}
\end{equation}
Unlike the case of $H\approx H_{s}$, this ``one-particle state'' cannot  be 
treated exactly, because the ``vacuum'' itself is not known exactly due to the 
non-integrablity of the system (except for some special values of ${\beta}$ 
($=\pm1,\infty$)). 
Accordingly, for general $\beta$, it is impossible to calculate the exact 
two-body $S$-matrix $S(k,k')$ which is utilized in the previous section to 
determine the effective coupling constant $c$.  
Nevertheless, if we assume the bose-gas picture to be held, we can 
``indirectly'' determine the value of $c$ from the $M-H$ curve obtained by the 
PWFRG.  
To see whether the obtained value of $c$ lies in a reasonable range or not, 
serves as a partial check of the validity of the bose-gas picture.

Near $H=H_{c}=\Delta$, we should relate the $M-H$ curve to the bose-gas energy 
density $\epsilon(\rho)$ as
\begin{eqnarray}
M=\rho,\nonumber\\
H-H_{c}=\frac{\partial\epsilon(\rho)}{\partial\rho}.\label{tag-18}
\end{eqnarray}
Then, for the square-root behavior
\begin{equation}
M \sim \sqrt{H-H_{c}} \quad(H\rightarrow H_{c}+0),\label{tag-19}
\end{equation}
we should expect the expansion of the form,
\begin{equation}
H=H_{c}+ 3A_{3}M^2 + 4A_{4}M^{3}+ 5A_{5}M^{4} +\cdots, \label{tag-20}
\end{equation}
where we have used (\ref{tag-3}).  
Since the expression (\ref{tag-17}) of the one-particle energy implies 
$\sigma=v^2/(2\Delta)$ in (\ref{tag-1}) and (\ref{tag-4}), we have
\begin{eqnarray}
A_{3}&=&\frac{\sigma\pi^2}{3}=\frac{\pi^2v^2}{6\Delta},\\
A_{4}&=&-4v^4\pi^2/(12\Delta^2 c). \label{tag-21}
\end{eqnarray}
From (\ref{tag-2})-(\ref{tag-4}), it is clear that the width of 
critical region essentially depends on the reduced coupling constant 
$\tilde{c}$ defined by 
(see (\ref{tag-4}))
\begin{eqnarray}
\tilde{c}&=&c/\sigma,\nonumber\\
         &=&-4A_{3}/A_{4}. \label{tag-22}
\end{eqnarray}
If we rewrite (\ref{tag-20}) as
\begin{equation}
H-H_{c}=\alpha\pi^2 M^2(1+\gamma M + \delta M^2) \quad(+O(M^5)),\label{tag-23}
\end{equation}
a condition for the square-root criticality is $|\gamma M| <<1$ 
(and also $|\delta M| <<1$). 
By $W_{M}^{(c)}$, we denote width of the critical region in $M$, which we 
conveniently define as
\begin{equation}
W_{M}^{(c)} =0.1/|\gamma|. \label{tag-24}
\end{equation}
Then $M<W_{M}^{(c)}$ implies $|\gamma M|<0.1$ which may be regarded as a 
necessary condition for the criticality.  
Correspondingly, we can introduce $W_{H}^{(c)}$ defined by
\begin{equation}
W_{H}^{(c)}=\alpha\pi^2 (W_{M}^{(c)})^2, \label{tag-25}
\end{equation}
which represents the width of the critical region in $H-H_{c}$.

Let us now check the bose-gas prediction of the $M-H$ curve by comparing it 
with the PWFRG calculations.  
From the $H-M$ curve, we determine $\Delta$ (=$H_{c}$), $\alpha$, $\gamma$ and 
$\delta$ in (\ref{tag-23}) by the least-square fitting. 
In Fig.3 we show the PWFRG results of the $H-M$ curves near $H_{c}$ 
for ${\beta}=0,1/3$. 
The obtained values of $\Delta$ are $0.410$ (for ${\beta}=0)$ and $0.699$ (for 
${\beta}=1/3$), both of which are in good agreement with the known values 
$0.4105$ (for the former~\cite{gap-heis}) and $0.699$ (for the 
latter~\cite{Mikeska,structure-factor}).

To verify the relation $\alpha=\sigma=v^2/(2\Delta)$ we need values of $v$.  
For ${\beta}=0$ using the known value~\cite{Sorensen-Affleck,dispersion} 
$v=2.46$ we have $\sigma=7.38$ which should be compared with $\alpha=7.65$ 
obtained from the $H-M$ curve; 
the obtained value of $\alpha$ is in reasonable agreement with $\sigma$.  
For ${\beta}=1/3$ there seems to be no serious numerical evaluation of $v$. 
We therefore consult Ref.\cite{Mikeska} where a variational calculation of 
$\omega(k)$ beyond the single-mode approximation~\cite{SMA} (which gives 
$\sigma=5/9=0.555\ldots$) is made; we have
\begin{eqnarray}
\sigma&=&(32645+359\sqrt{6529})/117522 \nonumber\\
      &=&0.5246\ldots .\label{tag-26}
\end{eqnarray}
This value is also in reasonable agreement with $\alpha=0.487$ obtained from 
the $H-M$ curve.  
Hence, the $H-M$ curves reproduce the ``one-particle quantities'' in the 
bose-gas picture. 
Further, the coefficients $\gamma$ and $\delta$ are estimated to be
\begin{eqnarray}
{\beta}=0:&& \gamma=-8.65,\quad \delta= 32.6\\
{\beta}=1/3:&& \gamma=-3.63,\quad \delta= 7.46  .\label{tag-27}
\end{eqnarray}
Since the negative values of $\gamma$ implies the positive effective 
bose-coupling constant, our PWFRG calculation supports the validity of the 
bose-gas picture for ${\beta}=0, 1/3$.

We should point out that, although the bose-gas prediction for the square-root 
behavior seems to be valid, the ``critical region'' 
of the square-root behavior in the $M-H$ curve is rather narrow, 
since the obtained values of $\gamma$ and $\delta$ are non-negligibly large.  
In fact, the quantity $W_{M}^{(c)}$ defined by (\ref{tag-24}) characterizing 
the width of the critical region, is very small: 0.012 (for ${\beta}=0$) and 
0.028 (for ${\beta}=1/3$).  
Corresponding values of $W_{H}^{(c)}$ defined by (\ref{tag-25}) are even 
smaller: $0.010$ (${\beta}=0$) and $3.8\times 10^{-3}$ (${\beta}=1/3$). 

  One notable behavior which we found in the PWFRG calculation is that, on 
raising ${\beta}$ from 0, the $H-M$ curve becomes flatter and flatter, or 
equivalently, the value of $\alpha$ in (\ref{tag-23}) becomes smaller and 
smaller; 
there seems to be a critical value ${\beta}_{c}$ ($\approx 0.41$) at which 
$\alpha$ vanishes.  
Accordingly, the critical behavior of the $M-H$ curve at  $H_{c}$ changes from 
square-root type to another one $\sim (H-H_{c})^{\theta}$ 
($\theta\approx0.25$) (Fig.4).  
In the bose-gas picture, this change of the $M-H$ curve may be understood as 
the vanishing of the $\bar{k}^2$-term in the expansion of the one-particle 
excitation energy $\omega(k)$.  
Interestingly, a qualitative change of the static structure factor $S(q)$ has 
been found~\cite{structure-factor} very near ${\beta}_{c}$.  
Since both of these changes reflect changes in the ground state and the 
low-energy excitation mode of the system, 
it is likely that they have a common origin.

Above ${\beta}_{c}$, the square-root behavior reappears. 
However, the coefficient $\gamma$ becomes positive (although small), implying 
{\em negative} effective coupling constant (Fig.5).  
The square-root behavior itself becomes manifest due to small $|\gamma|$, but 
the negative coupling disables us to take the naive bose-gas picture in this 
region of ${\beta}$; for justification of the picture, we should inspect the 
``bound states'', just as we did in discussing the $M-H$ curve near the upper 
critical field $H_{s}$.  
Note that for systems in the Haldane phase where the orientational order 
(characterized by the string order parameter~\cite{Rom-denNijs}) exists, the 
``particle" is a moving domain wall separating two regions each of which has 
complete orientational order.\cite{Tasaki,Mikeska,Fath-Solyom}  
In this view, total $S^{z}$ carried by a low-lying excitation mode is the 
``height'' of the wall.  
Then, if the wall width is narrow ($\sim$ one lattice spacing), we can adopt a 
similar reasoning as in the previous section justifying the negative coupling 
bose-gas picture: 
formation of stable bound states will be forbidden due to the kinematical 
constraint that the local wall height ($\sim$ total $S^{z}$, for thin wall) 
cannot exceed $S$.  
The actual situation is, however, subtle because the domain-wall is somewhat 
fuzzy (due to the zero-spin defects~\cite{Tasaki}) and its width may not be 
narrow.~\cite{Mikeska} 
In this view, we should say that full justification of the bose-gas picture 
for ${\beta}>{\beta}_{c}$ seems to require further study. 
Nevertheless, the square-root behavior $M\sim \sqrt{H-H_{c}}$ itself is 
confirmed by our PWFRG calculation.

\section{Summary}

In this paper, we have studied the zero-temperature magnetization process 
($M-H$ curve) of the $S=1$ isotropic antiferromagnetic spin chain 
with both the bilinear and biquadratic forms of interactions 
in the range $0\leq{\beta}<1$ 
where ${\beta}$ is the coefficient ratio between the biquadratic term and the 
bilinear term.  
Quantitative test for the bose-gas picture near the critical fields $H_{s}$ 
(saturation field) and $H_{c}$ (lower critical field) has been made with the 
help of the product-wavefunction renormalization-group (PWFRG) method which is 
a variant of S.R. White's density-matrix renormalization group (DMRG).

Near $H_{s}$ we have derived the correct effective bose-gas coupling constant 
from the two-down-spin scattering matrix in its low-energy limit. 
The resulting delta-function bose gas yields $M-H$ curves which are in good 
agreement with the PWFRG calculations. 

Near $H_{c}$, the square-root behavior $M\sim \sqrt{H-H_{c}}$ has been 
confirmed by our PWFRG calculation throughout the range of ${\beta}$ studied.
Here it should be noted a recent finite size scaling calculation by Sakai and 
Takahashi gave a consistent result for the $\beta=0$ 
case.\cite{Sakai-Takahashi2}
We have, however, found two distinct regions of ${\beta}$ separated by a 
critical value ${\beta}_{c}\approx 0.41$. In the small ${\beta}$ region, 
$0<{\beta}<{\beta}_{c}$, the effective bose-gas coupling $c$ extracted 
from the 
PWFRG-calculated $M-H$ curve is positive but small, making the critical region 
of the square-root behavior rather narrow; it becomes narrower and narrower on 
approaching ${\beta}_{c}$. At ${\beta}_{c}$, 
the $M-H$ curve seems to exhibit a 
different critical behavior $M\sim (H-H_c)^{\theta}$ with $\theta\approx0.25$. 
In the large ${\beta}$ region, although the square-root behavior is more 
pronounced due to large value of $|c|$, the sign of $c$ becomes negative, which 
sharply contrasts to the small-${\beta}$ region.

As regards the $M-H$ curve of the bilinear-biquadratic Heisenberg chain, 
cusp-like singularities in the ``middle-field'' region have been known for 
integrable ${\rm SU}(N)$ chains.\cite{SU(N)}  
Whether a similar behavior can also be found for general, non-integrable cases 
is an interesting problem.  
Although we have concentrated on the behavior near the critical fields in the 
present paper, we have obtained a full $M-H$ curve from $H=0$ to $H=H_{s}$. 
In the large ${\beta}$ region, we have actually found a clear cusp-like 
singularity very similar to the one in the $SU(3)$ (Lai-Sutherland) 
model,\cite{Lai-Sutherland} 
whose detailed account will be given in a separate paper.

Finally we would like to remark that the bose-gas description which we 
investigated in the present paper may not be the only one for ``quantitative'' 
description of the $M-H$ curve of the AF spin chain.  
For example, in a recent paper, Yamamoto\cite{Yamamoto} gave a different 
picture for the ground-state properties of the bilinear-biquadratic chain. 
Such an analysis may be helpful for clarifying nature of the system in the 
region ${\beta}>{\beta}_{c}$. 
Also, ``quantifying'' other low-energy effective theories is 
an interesting and important problem.  
For this purpose, the approach we have taken in section III where microscopic 
quantities of the effective theory are extracted from bulk quantities 
calculated by a reliable method, like the DMRG.

\acknowledgments

The authors would like to thank T. Nishino and H. Kiwata 
for valuable discussions.
This work was partially supported by the Grant-in-Aid for Scientific Research 
from Ministry of Education, Science, Sports and Culture (No.09640462).
K. O. is supported by JSPS fellowship for young scientists and Y. H. is partly 
supported by the Sasakawa Scientific Research Grant from The Japan Science 
Society.

\begin{figure}
\caption{Comparisons of the effective $\delta$-bose gas 
model and the PWFRG calculations near the saturation field $H_s$ for 
some values of the biquadratic interaction $\beta$.
(a) $\beta=0$, (b) $\beta=1/3$, (c) $\beta=0.6$.
The open circles represent the PWFRG results with the retained number 
of bases $m=60$.
The solid lines show the effective $\delta$-bose gas model with the coupling 
$c=-(3\beta+1)/\beta$.
In (b) and (c), we draw the free fermion curves, which correspond to $|c|=\infty$, as the broken lines for comparison.
}
\end{figure}

\begin{figure}
\caption{Comparisons of the effective $\delta$-bose gas 
model and the PWFRG calculations for various $S$ at $\beta=0$.
The diamonds, squares, triangles and circles represent the PWFRG results for 
$S=1/2$, $1$, $3/2$ and $2$ respectively.
The solid lines represent the effective $\delta$-bose gas model with 
$c=2/(S-1)$.
}
\end{figure}

\begin{figure}
\caption{Comparisons of the effective $\delta$-bose gas 
model and the PWFRG calculations near the lower critical field 
$H_c(=\Delta)$.
(a) The pure AF Heisenberg point $\beta=0$. (b) The AKLT point $\beta=1/3$.
The open circles represent the PWFRG results 
with the retained number of bases 
$m=100$.
The solid lines show the least-square fitting results of the form:
$H=\Delta+\alpha\pi^2 M^2(1+\gamma M + \delta M^2)$.
}
\end{figure}

\begin{figure}
\caption{The $H-M$ curve at $\beta=0.4(\approx \beta_c)$ near the lower 
critical field $H_c(=\Delta)$.
The open circles represent the PWFRG result with $m=100$.
The solid line shows the least-square fitting result of the form:
$H=\Delta+ A M^4 +B M^5$.
}
\end{figure}

\begin{figure}
\caption{The $H-M$ curve at $\beta=0.6$ near the lower critical field 
$H_c(=\Delta)$.
The open circles represent the PWFRG result with $m=100$.
The solid line shows the least-square fitting result.
We see the bose gas coupling constant takes the negative value.
}
\end{figure}

\end{document}